\documentclass[amsmath,amssymb,twocolumn]{revtex4-2}
\usepackage{graphicx}
\usepackage{subfig}
\usepackage{epsfig}
\usepackage{dcolumn}
\usepackage{multirow}
\usepackage{rotating}
\usepackage{verbatim}
\usepackage{bm}
\usepackage{color}
\usepackage{soul}
\usepackage{float}
\usepackage{hyperref}
\usepackage[utf8]{inputenc}

\def\lsim{\raise0.3ex\hbox{$<$\kern-0.75em\raise-1.1ex\hbox{$\sim$}}}
\def\gsim{\raise0.3ex\hbox{$>$\kern-0.75em\raise-1.1ex\hbox{$\sim$}}}

\def\beqa{\begin{eqnarray}}
\def\eeqa{\end{eqnarray}}

\begin{document}

\title{Precise determination of pomeron intercept via scaling entropy analysis}
\author{L.S. Moriggi$^{1}$}
\email{lucasmoriggi@unicentro.br}
\affiliation{$^{1}$ Universidade Estadual do Centro-Oeste (UNICENTRO), Campus Cedeteg, Guarapuava 85015-430, Brazil}
\author{M.V.T. Machado$^{2}$}
\email{magnus@if.ufrgs.br}
\affiliation{$^{2}$ High Energy Physics Phenomenology Group, GFPAE. Institute of Physics, Federal University of Rio Grande do Sul (UFRGS)\\
Caixa Postal 15051, CEP 91501-970, Porto Alegre, RS, Brazil} 

\begin{abstract}

In this work, we confront the geometrical scaling properties of inclusive DIS cross section ($e+p\rightarrow e +X$) with the scaling entropy obtained from event multiplicity. We show that these two quantities are equivalent in the kinematic range probed by H1 Collaboration data. 
We propose that scaling entropy associated with partonic interactions is a more efficient way to detect scaling in experimental data. We used a combined analysis of the inclusive cross section and entropy obtained from multiplicities $P(N)$ of final-state hadrons to accurately determine the value of the Pomeron intercept. The approach could provide new constraints for future hadron collider experiments and deepen our understanding of parton saturation.

\end{abstract}

\maketitle

%\tableofcontents

\section{Introduction}

Saturation physics predicts that at small Bjorken-$x$, the transverse momentum-dependent gluon distribution grows rapidly as $\phi(x,k_T) \sim x^{-\lambda}$, where $\alpha_P = \lambda + 1$ represents the hard Pomeron intercept. To prevent indefinite growth, non-linear effects become significant at the momentum scale $Q_s^2 \sim \phi(x) \sim x^{-\lambda}$. The gluon distributions are expected to depend on the scaling variable $k_T/Q_s(x)$ rather than on $x$ and $k_T$ independently, with this scaling behavior influencing various observables.

The scaling properties of inclusive cross sections at small-$x$ in deep inelastic scattering (DIS) are well established \cite{Golec-Biernat:1998zce,Golec-Biernat:1999qor,Stasto:2000er,Levin:1999mw,Iancu:2002tr,Praszalowicz:2012zh,Gelis:2006bs}. This phenomenon is one of the most compelling pieces of evidence for saturation physics. A similar scaling behavior has been observed at hadron colliders in the inclusive differential cross section for hadron production \cite{Moriggi:2020zbv,Osada:2019oor,McLerran:2014apa}. Studies show that the kinematic range of this scaling is consistent across DIS cross sections and hadron production, extending to values of momentum $Q^2$ significantly larger than $Q_s^2(x)$. By defining a scaling variable $\tau = Q^2/Q_s^2(x)$, it has been shown \cite{Moriggi:2020zbv} that the scaling behavior of inclusive cross sections in both $e + p \to e + X$ and $p + p \to h + X$ processes holds within the kinematical ranges $\tau < 10^3$ and $x < 0.08$.

The scaling properties observed in hadroproduction cross sections at small-$x$ are extended to multiplicity, geometry parameters, and hadron species in $pp$ and $pA$ collisions \cite{Moriggi:2024tbr,Osada:2020zui,Praszalowicz:2013fsa,McLerran:2014apa}. These properties serve as valuable tools for analyzing differential cross sections with dependencies on multiple parameters. However, detecting scaling in inclusive cross sections is model-dependent, requiring gluon distributions to be estimated within a specific framework and the parameter $\lambda$ to be extracted from experimental data. In the dipole picture of DIS, various models have been employed to extract $\lambda$ from the HERA combined data \cite{Rezaeian:2013tka,Iancu:2003ge,Golec-Biernat:2017lfv}, yielding different values (a comparison is presented in Section~\ref{sec:results}). Here, we propose that scaling entropy provides a model-independent method to determine $\lambda$ using event multiplicity data, provided the scaling hypothesis holds within a specific kinematic range.

Although not all QCD models predict scaling, $\lambda$ remains a crucial parameter to distinguish different parton dynamics. HERA data \cite{H1:1996zow,H1:1997uzt} have demonstrated the growth of the proton structure function $F_2(x,Q^2) \sim x^{-\lambda(Q^2)}$ at fixed values of $Q^2$, with $\lambda(Q^2)$ varying significantly from $\lambda(Q^2=1) = 0.12$ to $\lambda(Q^2=800) = 0.5$. At high $Q^2$, collinear factorization describes this growth by resumming powers of $\log(Q^2)$, while at low $x$, BFKL dynamics \cite{BFKL1,BFKL2} resums powers of $\log(1/x)$, yielding unintegrated gluon distributions with power-law growth $\phi(x,k_T) \sim x^{-\lambda}$ at a constant rate. Comparisons of these approaches in the description of the DIS data can be found in \cite{Ball:2017otu}. Furthermore, phenomenological models based on Regge theory, incorporating soft and hard Pomerons, have been used to fit the data \cite{ZEUS:2000sac,Abramowicz:1997ms,Abt:2017nkc}.

In recent years, the study of hadroproduction at high energies has evolved beyond collinear factorization and Regge phenomenology to address anomalous phenomena observed at the LHC. Collective behavior in small collision systems ($pp$, $pA$) with high multiplicities, such as strangeness enhancement \cite{ALICE:2016fzo,ALICE:2018pal,CMS:2016zzh}, angular correlations \cite{CMS:2010ifv,CMS:2015fgy}, and the multiplicity dependence of average $p_T$ \cite{ALICE:2018pal}, resemble observations in heavy-ion collisions. Theoretical frameworks have also been developed to describe the production and thermalization of quark-gluon plasma (QGP) in heavy-ion collisions \cite{Schenke:2012wb,Paatelainen:2012at,Kurkela:2015qoa,Baier:2000sb}.

Traditional Monte Carlo event generators such as PYTHIA \cite{Sjostrand:2006za,Sjostrand:2014zea} and EPOS LHC \cite{Pierog:2013ria} face challenges in accurately reproducing the multiplicity dependence of $p_T$-differential spectra in $pp$ collisions \cite{ALICE:2019dfi} and high-multiplicity $ep$ collisions \cite{H1:2020zpd}. In this context, information-theory and thermal descriptions of partonic degrees of freedom have gained traction \cite{Deppman:2016fxs,Moriggi:2024tbr,Peschanski:2012cw,Baker:2017wtt,Feal:2018ptp,Han:2018wsw,Chen:2024dhz,Tu:2019ouv,Kharzeev:2017qzs}. Entanglement entropy, proposed in \cite{Kharzeev:2017qzs}, has emerged as a valuable tool for testing QCD and has been applied to hadron production in DIS and small-$x$ parton cascades \cite{Hentschinski:2021aux,Hentschinski:2023izh,Hentschinski:2024gaa,Gotsman:2020bjc,Datta:2024hpn}. Entropy-based approaches provide new insights into QCD phenomenology, offering a means to control infrared divergences and study event multiplicity distributions.

This paper investigates scaling entropy as a model-independent tool to determine the parameter $\lambda$, offering a perspective on the dynamics of QCD at high energies.

The structure of the paper is as follows. Section~\ref{sec:model} introduces the scaling entropy as a consequence of unitarity and provides the formalism for partonic entropy in the Boltzmann-Gibbs form. The negative binomial distribution (NBD), used to extract experimental entropy, is also discussed. Section~\ref{sec:results} compares the entropy-derived $\lambda$ with the values obtained from inclusive cross-section scaling. Finally, the paper concludes with a summary of key findings.

\section{Theoretical framework and main predictions}
\label{sec:model}

At high energies, the proton wave function is dominated by a large number of soft gluons. The partonic content of protons is most effectively probed via the interaction of a virtual photon in deep inelastic scattering (DIS). A photon with virtuality $Q^2$ interacts with a proton characterized by a gluon distribution $\phi(x, k_T)$. This process is best understood in the color dipole picture, where the photon fluctuates into a quark-antiquark pair (a QCD color dipole). 

The key information on the proton structure comes from the imaginary part of the forward scattering amplitude of the dipole projectile with the proton target in the coordinate space, denoted $\mathcal{N}(x, r)$. A central feature of saturation physics is that the limited growth of the gluon distribution is linked to the unitarity constraint on the scattering amplitude, i.e. $\mathcal{N}(x, r)$ is bounded between 0 (dilute regime) and 1 (saturation regime). In dipole coordinate space, $r$, the cross section can be expressed as a convolution of the photon wavefunction $|\psi_{\gamma}(r,z)|^2$ with $\mathcal{N}(x, r)$:
\begin{equation} \label{eq:gammap}
    \sigma_{\gamma^* p}(x, Q^2) = \sigma_0 \int d^2r \, dz \, |\psi_{\gamma}(r,z)|^2 \, \mathcal{N}(x, r).
\end{equation}

This expression can be transformed into transverse momentum space via a Fourier transform:
\begin{equation} 
    \sigma_{\gamma^* p}(x, Q^2) = \sigma_0 \int d^2k_T \, dz \, |\psi_{\gamma}(k_T, z)|^2 \, \mathcal{P}(x, k_T),
\end{equation}
where the Fourier transform of the scattering amplitude, $\mathcal{P}(x, k_T)$, is normalized to unity and can be interpreted as a probability distribution containing all information about the interaction process at the partonic level. 

The form of $\mathcal{P}(x, k_T)$ can be derived from the maximum entropy principle using Lagrange multipliers to optimize the distribution in a canonical ensemble. The most general form of nonextensive statistical mechanics, the Tsallis entropy \cite{Tsallis:1987eu}, is given by:
\begin{equation}\label{eq:tsallis}
    S_q = \int d^2k_T \frac{1 - [\mathcal{P}(x, k_T)]^q}{q - 1},
\end{equation}
with constraints:
\begin{equation} \label{eq:ktm}
    \langle k_T^2 \rangle_q = \frac{\int d^2k_T k_T^2 [\mathcal{P}(k_T)]^q}{\int d^2k_T [\mathcal{P}(k_T)]^q} = \beta^{-1}, \quad \int d^2k_T \mathcal{P}(k_T) = 1.
\end{equation}
The Tsallis q-index represents the degree of non-extensivity of the distribution. 

This framework leads to a stationary state with a power-law distribution, as proposed in Refs. \cite{Moriggi:2020zbv, Moriggi:2024tbr}. The Lagrange parameter $\beta$ can be linked to the scaling hypothesis:
\begin{equation}\label{eq:einstein}
    \langle k_T^2(x) \rangle_q \sim \beta^{-1} (x_s/x)^{\lambda}.
\end{equation}

Since $1/x$ corresponds to the time scale probed by the soft gluons, the interaction with the gluons of varying energies results in an anomalous diffusion-like process. This connects $\langle k_T^2(x) \rangle_q$ to a generalized form of Einstein's relation for diffusion.

Under these conditions, the probability distribution can be expressed in a scaling form:
\begin{equation}\label{eq:scaling}
    \mathcal{P}(x, k_T^2) \sim \frac{1}{x^{-\lambda}} f(k_T^2 / x^{-\lambda}).
\end{equation}

The resulting partonic entropy in the Tsallis formalism is given by:
\begin{equation} \label{eq:entroQ}
    S_q^{\text{parton}}(x) = \frac{1}{q-1} - \left( \frac{2-q}{q-1} \right)^q (\pi Q_s^2(x))^{1-q}.
\end{equation}

This general form has been used to establish a relationship between the saturation scale and the overlap area in $pp$ collisions \cite{Moriggi:2024tbr}. Scaling exists regardless of whether $q = 1$ or $q \neq 1$; however, integrated $k_T$ data cannot distinguish differences associated with high-$k_T$ degrees of freedom. Such effects are observable in $p_T$-differential cross sections.

\begin{figure*}[t]
\includegraphics[width=0.8\linewidth]{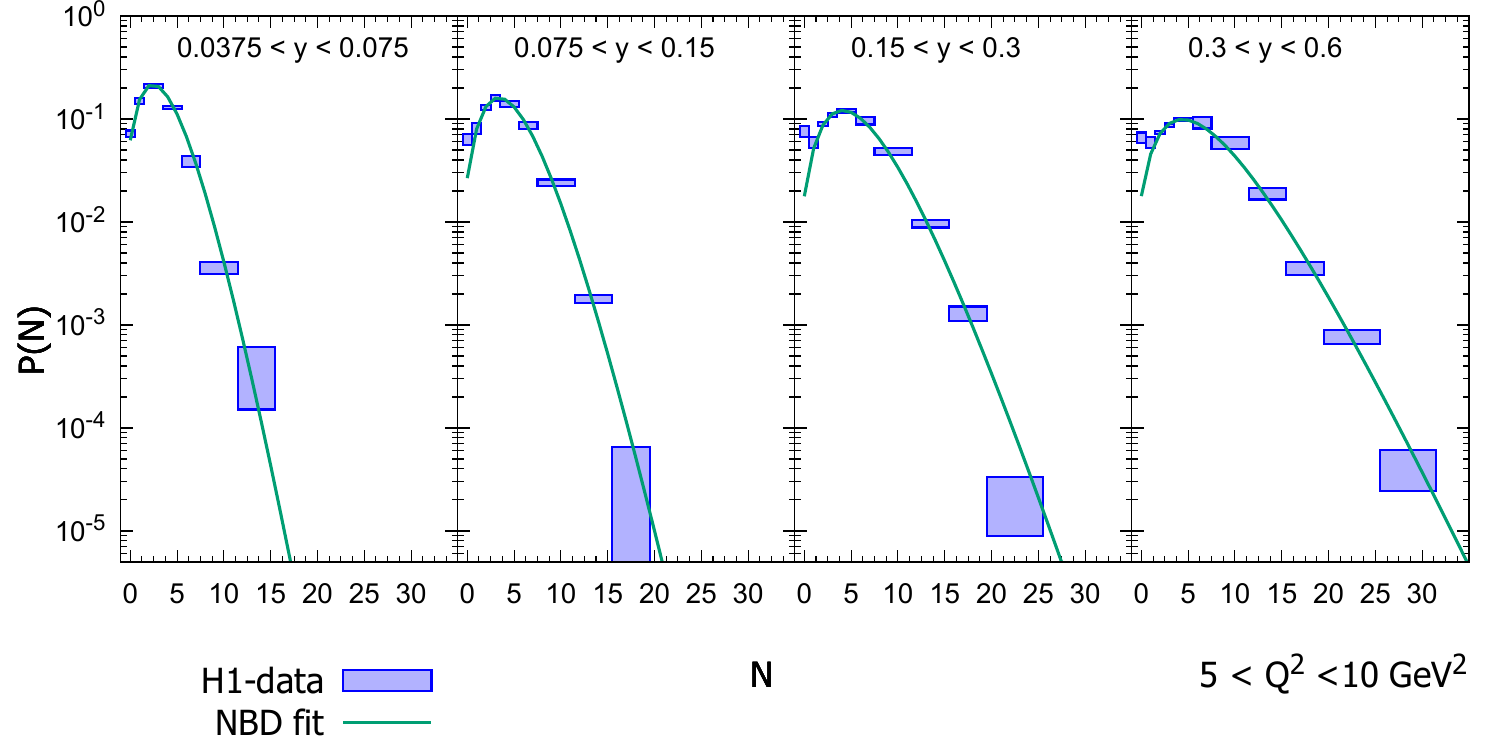}
\caption{The NBD multiplicity fit using Eq. \eqref{eq:nbd} at different values of $y$ (solid lines) compared with experimental data from H1 collaboration \cite{H1:2020zpd} (blue boxes). The fit for $P(N)$ versus $N$ is perform for  $N>1$.}\label{fig:multiplicity}
\end{figure*}

Experimentally determined entropy from multiplicity production is usually defined in the Boltzmann-Gibbs (BG) form and can also be estimated in $q=1$ limit:
\begin{equation}
    S_1^{\text{parton}}(x) = -\int \mathcal{P}(x, k_T) \log[\mathcal{P}(x, k_T)] d^2k_T.
\end{equation}

Assuming the scaling relation holds, the entropy results in the following form:
\begin{equation}\label{eq:pentro}
    S_1^{\text{parton}} = C + \lambda \log\left(\frac{1}{x}\right),
\end{equation}
where $C$ is a constant dependent on the specific form of $\mathcal{P}(x, k_T)$, but the slope with respect to $\log(1/x)$ directly provides $\lambda$. This scaling relation underpins the anomalous diffusion-like behavior of gluons and connects the entropy to partonic momentum distributions.

In this expression, the Pomeron intercept can be extracted experimentally from hadron multiplicity data, provided that scaling survives the hadronization process. According to the local-hadron-parton duality (LHPD) \cite{Dokshitzer:1987nm}, the hadron spectra reflect the partonic spectra, without an additional $x$-dependence introduced during hadronization. This is supported by scaling line analyses in $pp$ collisions \cite{Moriggi:2020zbv, Moriggi:2024tbr}, where the energy dependence (or dependence on $x$) of $p_T$-spectra remains consistent post-hadronization.

If we consider the power-like gluon distribution proposed in \cite{Moriggi:2020zbv} characterized by a power law parameter $2+\delta n$, the constant $C$ in the corresponding expression can be estimated as $C= \frac{(2+\delta n)}{(1+\delta n)} + \log\left(\frac{\pi}{1+\delta n}\right )+\lambda \log(x_0)\approx - 0.82$. The hadronization process results in an increase in the entropy of hadrons compared to that of partons. Specifically, if an ejected parton with momentum $Q$ gives rise to a hadron with momentum $\langle z \rangle Q$, we expect a corresponding increase in entropy given by
$ S^{hadron} = S^{parton} + 2\log\left(\frac{1}{\langle z \rangle}\right)$, which can have a dependence on $Q^2$. For comparison, taking $\langle z \rangle = 0.5$ yields $C \approx 0.57$.

The Boltzmann-Gibbs entropy is additive in parton rapidity ($Y=\log(1/x)$), i.e., $S^{\text{parton}}(Y_0 + \Delta Y) = S(Y_0) + S(\Delta Y)$, a property that breaks down when $q \neq 1$. Nonextensive characteristics of hadroproduction, reported in several works \cite{Wong:2015mba, Bhattacharyya:2016lrk, Bhattacharyya:2017cdk, Biro:2017arf, Biro:2020kve, Akhil:2023xpb, Li:2020lww, Sharma:2018jqf, Khuntia:2017ite, Parvan:2016rln}, suggest that this behavior applies to gluons under LHPD as well. However, for consistency, we must use the BG scaling entropy for comparison with experimental data as explained next.

We will now discuss the experimental aspects of extracting the experimental entropy from the HERA $ep$ data.
In multiplicity data, the entropy indicator related to the probability of detecting $N$ charged hadrons, $P(N)$, is investigated in BG form:
\begin{equation}\label{eq:entroexp}
    S^{\text{mult}} = -\sum_N P(N) \log(P(N)).
\end{equation}

The most established parametric model for describing multiplicities is the negative binomial distribution (NBD). In $pp$ collisions at LHC energies, a two-component NBD is often needed: a soft component for lower multiplicity events and a hard component for high-multiplicity events. However, for H1 data, a single component NBD is sufficient, as will be discussed in the section ~\ref{sec:results}. The NBD probability is given by
\begin{equation}\label{eq:nbd}
    P_{\text{NBD}}(N, \langle N \rangle, k) = \alpha \frac{\Gamma(N + k)}{\Gamma(N + 1) \Gamma(k)} \frac{(\langle N \rangle / k)^N}{(1 + \langle N \rangle / k)^{N + k}}.
\end{equation}
Here, $\langle N \rangle$ and $k$ are fitted parameters. Since NBD does not describe $N = 0$, data fit excludes $N = 0$, and $\alpha$ ensures normalization. 

The dispersion $D$ is given by:
\begin{equation}\label{eq:var}
    D^2 = \langle N^2 \rangle - \langle N \rangle^2 = \langle N \rangle + \langle N \rangle^2 / k.
\end{equation}

In DIS, Bjorken-$x$ is related to experimental quantities:
\begin{equation}
    x_{\text{Bj}} = \frac{Q^2}{sy}, \quad W = \sqrt{Q^2 \left(\frac{1}{x} - 1\right) + m_p^2},
\end{equation}
where $y$ is the fractional energy loss of the incoming electron, and $W$ is the hadronic center-of-mass energy. High $W$ corresponds to large invariant masses of final-state hadrons, associated with high-multiplicity events.

The main point of this work is a confrontation of parton scaling entropy \eqref{eq:pentro} with an experimentally defined entropy indicator from \eqref{eq:entroexp} obtained from NBD fit to multiplicity \eqref{eq:nbd}. In this way we are able to extract $\lambda$ in model independent way and compare with the inclusive cross section $\lambda$.

\begin{table*}[t]
\centering
\begin{tabular}{|c|c|c|c|c|c|}
\hline
$Q^2$ Range & $y$ Range & $\alpha$ & $\langle N \rangle$ & $k$ & $\chi^2/dof$ \\ \hline
$5 < Q^2 < 10$ & $0.0375 < y < 0.075$ & $1.050 \pm 0.033$ & $3.223 \pm 0.098$ & $11.02 \pm 3.053$ & 0.125 \\ 
               & $0.075 < y < 0.15$   & $0.952 \pm 0.017$ & $4.304 \pm 0.070$ & $9.938 \pm 1.029$ & 0.048 \\ 
               & $0.15 < y < 0.3$     & $0.903 \pm 0.020$ & $5.475 \pm 0.105$ & $6.225 \pm 0.485$ & 0.103 \\ 
               & $0.3 < y < 0.6$      & $0.886 \pm 0.025$ & $6.380 \pm 0.172$ & $4.281 \pm 0.334$ & 0.236 \\ \hline
$10 < Q^2 < 20$ & $0.0375 < y < 0.075$ & $1.019 \pm 0.016$ & $3.199 \pm 0.043$ & $40.98 \pm 18.53$ & 0.018 \\ 
                & $0.075 < y < 0.15$   & $0.972 \pm 0.017$ & $4.406 \pm 0.068$ & $10.92 \pm 1.146$ & 0.046 \\ 
                & $0.15 < y < 0.3$     & $0.922 \pm 0.021$ & $5.401 \pm 0.108$ & $6.071 \pm 0.493$ & 0.115 \\ 
                & $0.3 < y < 0.6$      & $0.889 \pm 0.022$ & $6.378 \pm 0.146$ & $4.539 \pm 0.311$ & 0.180 \\ \hline
$20 < Q^2 < 40$ & $0.0375 < y < 0.075$ & $1.016 \pm 0.019$ & $3.194 \pm 0.052$ & $18.87 \pm 5.221$ & 0.025 \\ 
                & $0.075 < y < 0.15$   & $1.007 \pm 0.012$ & $4.554 \pm 0.042$ & $12.41 \pm 0.865$ & 0.018 \\ 
                & $0.15 < y < 0.3$     & $0.962 \pm 0.020$ & $5.676 \pm 0.107$ & $7.537 \pm 0.615$ & 0.100 \\ 
                & $0.3 < y < 0.6$      & $0.911 \pm 0.020$ & $6.614 \pm 0.129$ & $4.790 \pm 0.291$ & 0.129 \\ \hline
$40 < Q^2 < 100$ & $0.0375 < y < 0.075$ & $0.967 \pm 0.053$ & $2.830 \pm 0.146$ & $12.68 \pm 8.785$ & 0.262 \\ 
                 & $0.075 < y < 0.15$   & $0.984 \pm 0.019$ & $4.566 \pm 0.067$ & $16.93 \pm 2.374$ & 0.047 \\ 
                 & $0.15 < y < 0.3$     & $0.989 \pm 0.018$ & $5.911 \pm 0.091$ & $7.814 \pm 0.538$ & 0.065 \\ 
                 & $0.3 < y < 0.6$      & $0.952 \pm 0.032$ & $6.641 \pm 0.174$ & $5.743 \pm 0.529$ & 0.237 \\ \hline
\end{tabular}
\caption{The resulting parameters for NBD distribution fit using  Eq. \eqref{eq:nbd} at different bins of $Q^2$ and $y$.}
\label{tab:nbd}
\end{table*}

\section{Results and discussions}
\label{sec:results}

We now proceed to the analysis of multiplicity data $P(N)$ in the context of NBD with one component. We observe that in the range of multiplicities $N<40$ for charged hadrons the second component as used in high energy hadronic collisions \cite{ALICE:2015olq,ALICE:2010mty,Grosse-Oetringhaus:2009eis} is not needed. We also exclude the $N=0$ bin that cannot be accurately described by NBD. We must remember that NBD is characterized by two parameters $\langle N \rangle$ and $k$ related to average and fluctuations that can be addressed in different theoretical models \cite{Giovannini:1985mz,Grosse-Oetringhaus:2009eis,Zborovsky:2018vyh,Gelis:2009wh,Gotsman:2020bjc}. Therefore, in this sense, it provides us with a general parameterization of the fluctuations in each event. 

The H1 collaboration extracted entropy and variance from the data using cubic spline interpolation between experimental points. Since we are using NBD fit for interpolation/extrapolation there is a small difference between the two approaches in those quantities. Regarding data binning at large multiplicity, we must consider the large uncertainties in the average $N$ in the larger bins. To account for this uncertainty we fit by minimization of $\chi^2$  using the effective variance,
\begin{equation}
\chi^2 = \sum_i \frac{(y_i-f(x))^2}{\sigma_{iy}^2+\left( \sigma_{ix} \frac{\partial f(x) }{\partial x} \right) ^2 }.
\end{equation}

In Figure \ref{fig:multiplicity} we present the comparison of data for $P(N)$ and the fit of the NBD function at different values of $y$. Data correspond to photon virtualities $5<Q^2<10$ GeV$^2$.  
The resulting parameters are presented in Table \ref{tab:nbd}. The parameter $1/k$ associated with variance \eqref{eq:var} shows the deviation from the Poisson distribution at $1/k \rightarrow 0$. At high $W$ ( larger $x$) we have $P(N)$ closer to the Poisson limit, and the variance grows fast up to $1/k \sim 0.2$ which is the value obtained from $pp$ colliders at $\sqrt{s} \sim 100 $ GeV \cite{Grosse-Oetringhaus:2009eis}.

An important aspect of hadroproduction is the insufficiency of one component NBD to describe $N \gsim 40 $. This situation is observed in LHC $pp$ multiplicity data \cite{ALICE:2015olq,ALICE:2010mty}. It is usually associated with a secondary mechanism of particle production and two component NBD (soft + hard) is used to fit $P(N)$ at high $N$ values. Although the small $x$ (higher $\langle N \rangle $) H1 data present a small deviation of one component of the NBD, it is not statistical sufficient to account for the second distribution.

Given that we have control over the probability distribution, we can now proceed with the entropy analysis. The experimental estimation of entropy can be obtained from the NBD distribution using Eq. \eqref{eq:entroexp}. Under the assumptions mentioned before, we can compare it with partonic entropy from Eq. \eqref{eq:pentro}. This allows us to extract the value of $\lambda$ to each value of $Q^2$ directly from entropy. 

\begin{figure*}[t]
\includegraphics[width=0.8\linewidth]{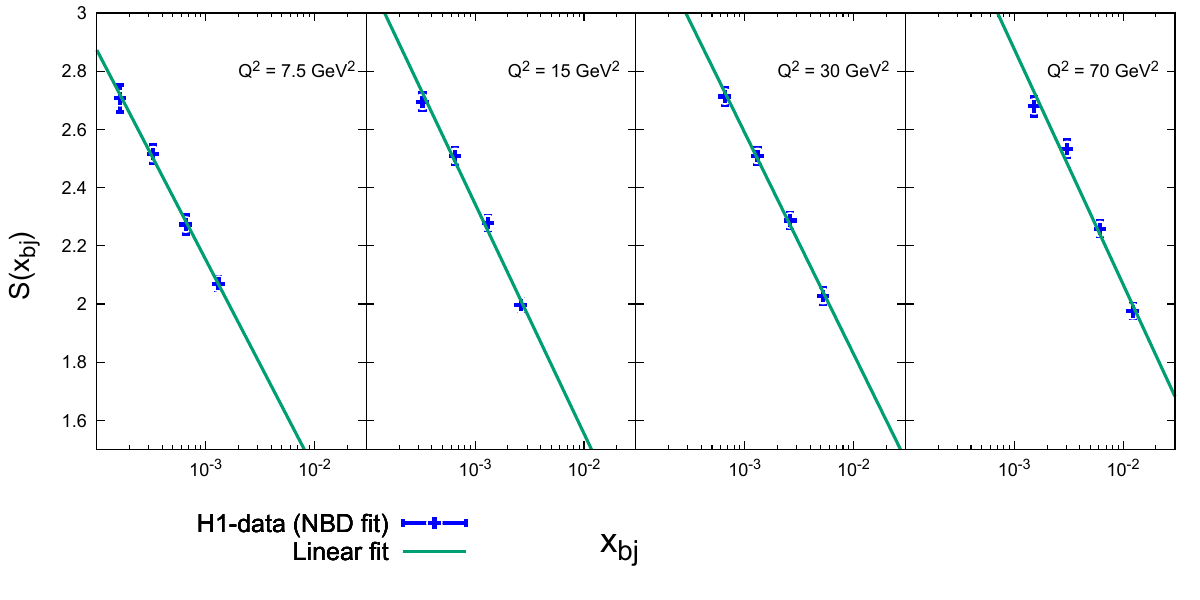}
\caption{The experimental entropy obtained from H1 data (bars) using the relation \eqref{eq:entroexp} with NBD extrapolation to high values of $N$. The solid lines represent the predicted scaling entropy given by Eq. \eqref{eq:pentro}. }\label{fig:entropy}
\end{figure*}

 Figure \ref{fig:entropy} shows the resulting linear behavior of partonic entropy in solid lines compared to the experimental value extracted from the estimation of NBD as a function of $x_{bj}$. There is a good agreement with the data in the range of multiplicities observed in the HERA kinematic range. There is no significant evidence in this range of high-order corrections like $\log^n(1/x)$ to the linear behavior of entropy. 

\begin{figure}[t]
\includegraphics[width=1.0\linewidth]{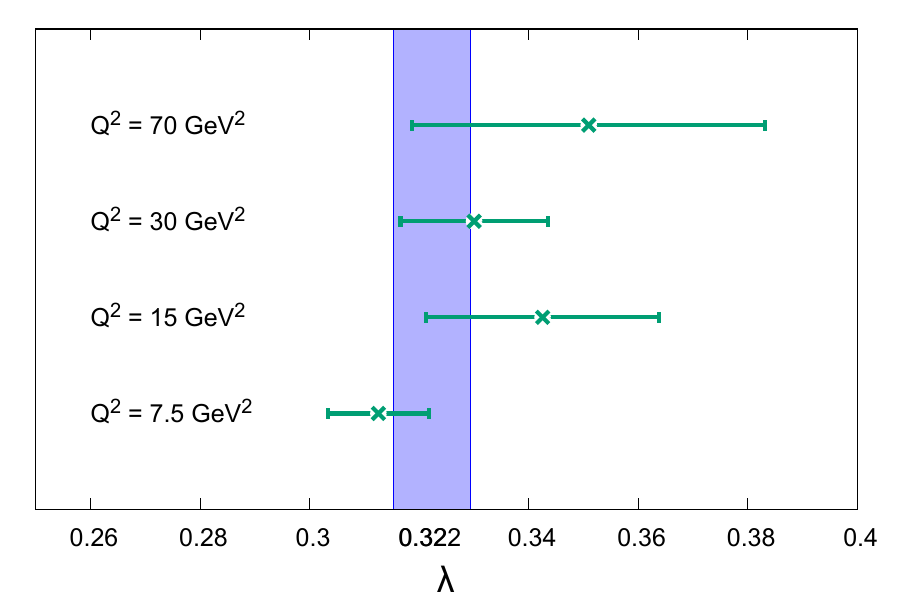}
\caption{Values of $\lambda$ obtained at each $Q^2$ bin (green bars). They are on average independent of $Q^2$ with the average weighted value plus uncertainty represented by the filled blue 
 vertical bar.}\label{fig:lambda}
\end{figure}

The values of the extracted values of $\lambda$ are presented in Figure \ref{fig:lambda}. The filled bar represents the average value with uncertainties from the different $Q^2$ bins. The final value obtained by the scaling of the partonic entropy is 
\begin{equation}\label{eq:eresult}
    \lambda_{entropy} = 0.322 \pm 0.007.
\end{equation}

Different works studied the scaling of the cross section in HERA and LHC with scaling models based on saturation physics. In order our hypothesis to make sense, i.e. that the quantity from Eq. \eqref{eq:pentro} can be used to describe the small-$x$ behavior of gluon distribution, the $\lambda$ value found in scaling of the inclusive cross section must be the same as found in entropy analysis.

Now we are going to confront the entropy scaling with inclusive cross-section scaling, $\sigma_{\gamma p}(\tau)$, in order to check the consistency of both descriptions.  Provided that the gluon distribution has the scaling property i.e. it can be written as a function of the variable $k_T/Q_s(x)$, for any particular form of the distribution the partonic entropy is the same except by some constant. So in this sense the scaling entropy is a model-independent way to determine the Pomeron intercept in scaling models. 

On the other hand the determination of scaling in cross-section data is dependent of the particular model for gluon distribution. We follow the analysis from \cite{Moriggi:2020zbv} where the gluon distribution is parametrized by a power law form in order to describe scaling in inclusive pion production cross section from hadron collider data. In that case, the value $\lambda = 0.33$ was used in order to reproduce the pion spectra scaling. 

\begin{figure}[t]
\includegraphics[width=1.0\linewidth]{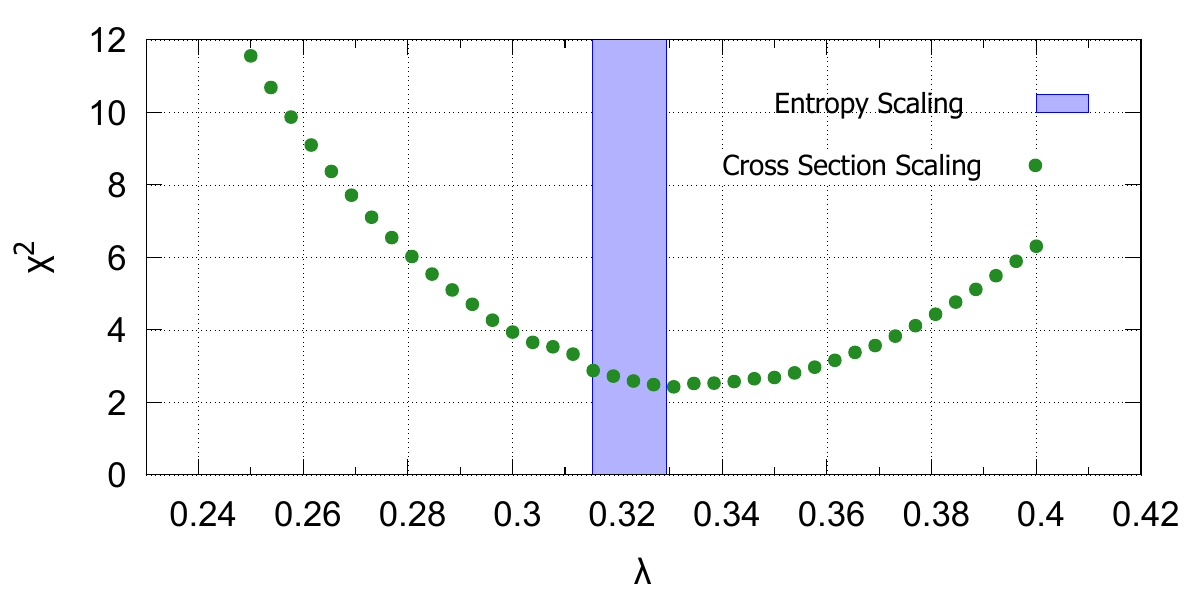}
\caption{The value of $\chi^2$ for the inclusive cross section fit as a function of different values of $\lambda$ (green dots) compared with scaling entropy extracted parameter (filled blue bar).}\label{fig:chi2}
\end{figure}

\begin{figure}[t]
\includegraphics[width=1.0\linewidth]{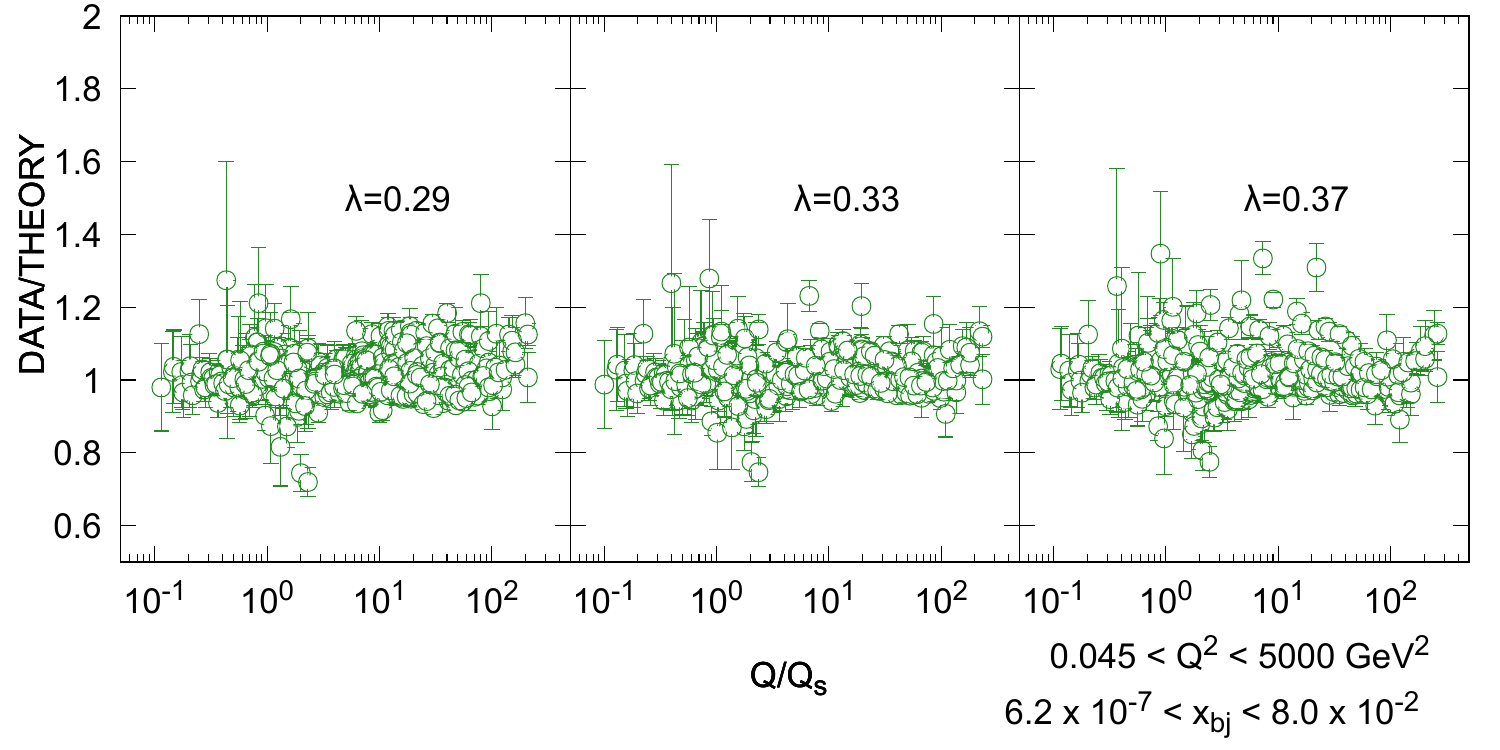}
\caption{Data to theory ratio between the MPM scaling model and combined HERA data at different values of $\lambda$. The central value is the one obtained from minimum $\chi^2$.}\label{fig:ratio}
\end{figure}

We now let $\lambda$ be a free parameter in the fit to the model. The $\gamma p $ cross section can be calculated in the dipole picture \eqref{eq:gammap} including only light flavors with zero quark mass. The resulting $\chi^2$ for each value of $\lambda$ is presented in Figure. \ref{fig:chi2} and compared with the result from scaling entropy. The minimum of the $\chi^2$ minimization curve coincides inside the uncertainty with the value of $\lambda_{entropy}$ indicating that this is a universal quantity that determines the partonic dynamics in the small-$x$ regime. The estimation of $\lambda$ by the inclusive cross section method is
\begin{equation}\label{eq:sresult}
    \lambda_{\sigma} = 0.329 \pm 0.025,
\end{equation}
which is in good agreement with Eq. \eqref{eq:eresult}. It is important to note that all data on the entropy are within the kinematic interval $x$ and $Q^2$ from inclusive cross section data where scaling of the cross section in terms of the variable $Q/Q_s(x)$ as shown by Fig. \ref{fig:ratio} is expected to be obeyed. 

It is timely to compare our results with different approaches in the literature. First, concerning the saturation models the determination of energy growth of saturation scale $Q^2_s(x)\sim x^{-\lambda}$ the leading logarithm (LL) prediction from BK equation gives $\lambda_{LL} = 4.88 N_c\alpha_s/\pi$. This produces a growth too fast for any phenomenological application. The NLO correction predicts slower growth $\lambda_{BK} \sim 0.3$ \cite{Albacete:2004gw}. On the other hand, models with impact parameter dependence usually give a lower $\lambda$ due to the growth of the proton radius that contributes to cross-section growth. The energy dependence of saturation scale in these models was estimated in \cite{Lappi:2011gu} using the IPsat model \cite{Kowalski:2003hm} resulting in the value $\lambda = 0.2$ and the bCGC model \cite{Watt:2007nr} which gives $\lambda \simeq 0.18$. 

\begin{figure}[t]
\includegraphics[width=1.0\linewidth]{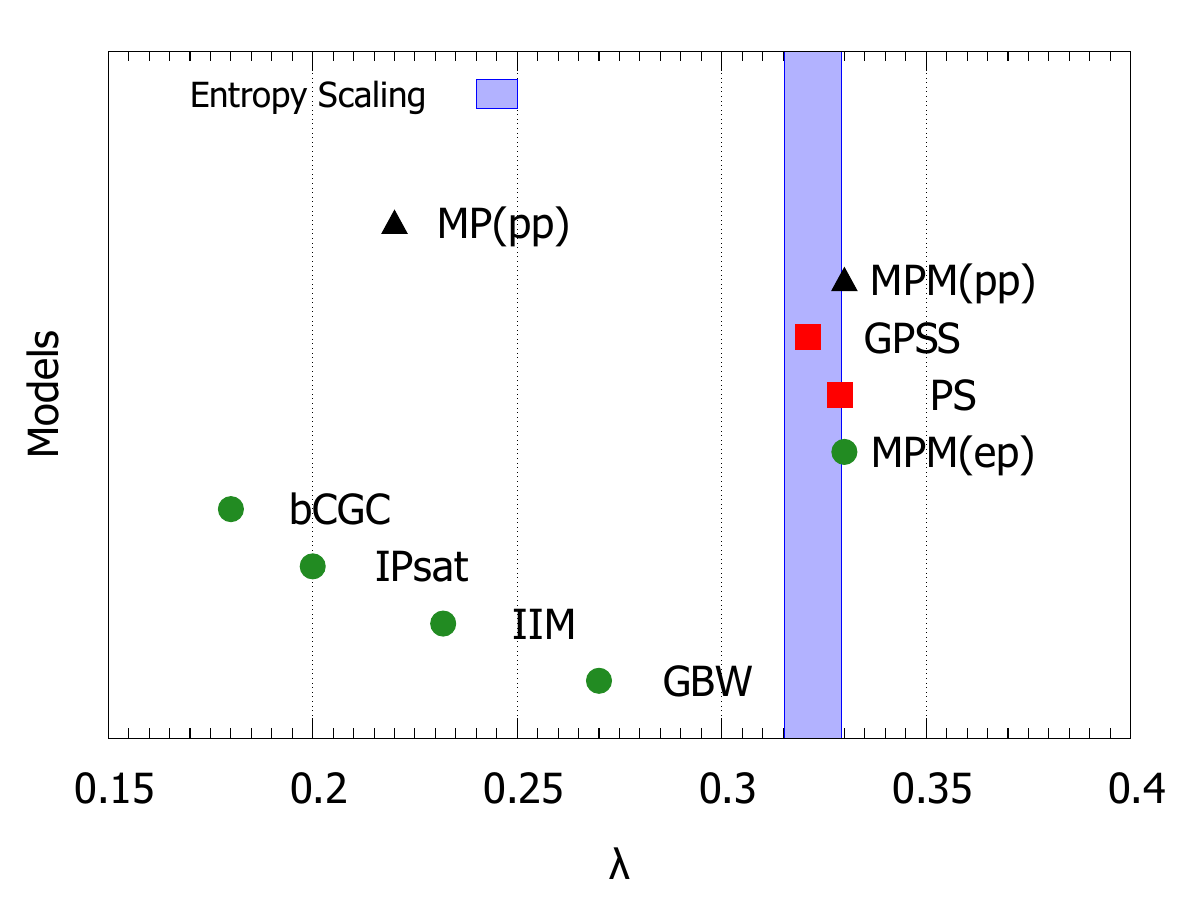}
\caption{Comparison of different values of $\lambda$ found at different works. See the discussion in the text.}\label{fig:models}
\end{figure}

Different dipole models used inclusive cross section $\gamma^* p$ to make the determination of $\lambda$. The values of $\lambda$ obtained from these models are presented in green circles in Figure \ref{fig:models}. The values extracted from impact parameter dependent models IPsat and bCGC are the ones estimated in Ref. \cite{Lappi:2011gu}.  Homogeneous impact parameter models such as GBW \cite{Golec-Biernat:2017lfv} and IIM \cite{Iancu:2003ge} lead to higher values close to the scaling line. However, they all present slower growth than that obtained by entropy scaling analysis. This is in part due to the kinematic range involved in fit, the number of fitted parameters, and also the uncertainties due to the gluon distribution modeling itself. On the other hand, model independent analysis using some quality factor indicator in a model independent way as done in Refs. \cite{Praszalowicz:2012zh} (PS) and \cite{Gelis:2006bs} (GPSS) present a larger value of $\lambda \sim 0.32 - 0.33$ compatible with the entropy scaling analysis.  The value obtained using the MPM model for $\sigma_{\gamma* p}$ is also presented in Figure \ref{fig:models} and is denoted by MPM(ep).

From the collision system $pp$, the MPM model \cite{Moriggi:2020zbv} uses the value $\lambda = 0.33$ in   order to reproduce scaling line for pion production inclusive cross section $pp(\bar p) \rightarrow \pi + X$ in $\sqrt{s}=200$ GeV (RHIC) up to $\sqrt{s}=13$ TeV (LHC). The value is different from the one considered in the analysis of Refs. \cite{McLerran:2014apa} (MP) and \cite{Osada:2019oor} using the charged hadron multiplicity-spectra at LHC energies. This is due to the fact that the authors used the number of charged hadrons and not the inclusive cross section for calculation of the scaling line as was done in the MPM case (denoted MPM(pp) in Fig. \ref{fig:models}). 

In conclusion, any model-independent analysis of geometric scaling of inclusive cross section in HERA data agrees with the scaling entropy scaling presented in this work. Its important to note that this is the same value observed also in inclusive cross section at hadron colliders. We hope that this analysis can help us understand the energy behavior of the saturation scale and make dipole models more accurate.

We now move to the discussion of the relationship between scaling entropy and other types of entropy proposed to deal with the phenomenology of QCD. In the model proposed by Kharzeev and Levin in Ref. \cite{Kharzeev:2017qzs} the entropic indicator of partonic degrees of freedom is associated with the integrated collinear gluon distribution function 
\begin{equation} \label{eq:KL}
    S_{KL} = \log(xG(x,Q^2)).
\end{equation}
The report from H1 collaboration \cite{H1:2020zpd} compares this indicator with experimental data showing large deviations from data/theory. We note that the result will be dependent of the model used to the collinear gluon distribution function. It will also generate running of the $\lambda(Q^2)$ given the dependence of $Q^2$ on the gluon distribution.  This quantity will be the same as the scaling entropy \eqref{eq:pentro} only if $xG(x,Q^2) = x^{-\lambda}$ is independent of $Q^2$. A second analysis presented in Ref.  \cite{Kharzeev:2021yyf} included a factor $C(\alpha_s, \log(Q^2), \log(1/x))$ in order to describe the splitting of virtual quark which has been introduced to correctly describe the H1 data. Along the same lines, a study is available in Ref. \cite{Hentschinski:2022rsa} confront different approaches to the gluon distribution with H1 data. This approach has been used in phenomenological applications such as \cite{Datta:2024hpn} to study the hadronization process and multiplicities in $pp$ collision at LHC \cite{Tu:2019ouv}. Moreover, a more accurate description is provided in Ref.  \cite{Hentschinski:2021aux} has tested this approach using a modification of relation \eqref{eq:KL} to include (sea) quarks, and the results show agreement with H1 data by using updated collinear PDFs.

The resulting ratio data/theory for $\sigma^{\gamma p }$ is presented to different values of $\lambda$ in Figure \ref{fig:ratio}. All cross-section data over a wide range of variables $x$, $Q^2$ are in good agreement close to the scaling line in relation to the variable $Q/Q_s(x)$. Although the parameter $\lambda$ has been extracted from the inclusive cross-section of HERA data in different works, particle production at the LHC has highlighted the importance of this scaling for understanding hadronic collision physics. However, determining this scaling can be model-dependent or difficult to observe directly in the data, and scaling entropy may prove useful for detecting scaling in such cases.

The detection of particles at the LHC has revealed the challenge of dealing with differential cross-sections that depend on multiple variables, such as impact parameters, multiplicity classes, fluctuations, centrality, rapidity, energy, and momentum scales involved in the collision, among others. In these scenarios, the idea presented here is expected to be of practical utility.

%\begin{table*}[ht]
%\centering
%\begin{tabular}{|c|c|c|c|}
%\hline
%\textbf{$Q^2$ (GeV$^2$)} & \textbf{$\lambda$ ($\pm$ Uncertainty)} & %\textbf{cte ($\pm$ Uncertainty)} & \textbf{$\chi^2/dof$} \\ \hline
%7.5           & $0.312568 \pm 0.009$             & $-0.006527 \pm %0.0677$            & 0.192787         \\ \hline
%15           & $0.342548 \pm 0.0212$             & $-0.022968 \pm %0.1446$            & 1.878396         \\ \hline
%30           & $0.330039 \pm 0.0134$             & $0.308535 \pm %0.0844$             & 0.562880         \\ \hline
%70           & $0.350979 \pm 0.0322$             & $0.451385 \pm %0.1720$             & 3.226340         \\ \hline
%\textbf{Average}          & $0.3223 \pm 0.007$             & -        %                            & -                \\ \hline
%\end{tabular}
%\caption{$\lambda$ values obtained from each $Q^2$ bin via scaling %entropy fit given by Eq. \ref{eq:pentro}.}
%\label{tab:lambda}
%\end{table*}

\section{SUMMARY AND CONCLUSIONS} \label{sec:conclusion}

We present an entropic indicator that can be obtained in a model-independent way from experimental data relating partonic degrees of freedom with final state hadrons. We show the utility of such concept by analyzing H1 data on charged-hadron multiplicities and use it to extract the Pomeron intercept.  We conclude that the value obtained from scaling entropy is the same as obtained from inclusive cross-section data indicating robustness of such approach on determining important QCD quantities.  The scaling entropy concept can also be used to detect scaling in different experimental data in a direct way without direct calculation of cross sections. In future works the same approach can be used to confront this entropic indicator with cross section to different processes at LHC.

\section*{Acknowledgments}

MVTM acknowledges funding from the Brazilian agency Conselho Nacional de Desenvolvimento Cient\'ifico e Tecnol\'ogico (CNPq) with the grant CNPq/303075/2022-8.

\bibliographystyle{h-physrev}
\bibliography{scalingentropy}

\end{document}